\documentclass{article}
\usepackage[T1]{fontenc} 
\usepackage[utf8]{inputenc} 
\usepackage{amsmath,cite,url}
\usepackage{graphicx}
\usepackage{color}
\usepackage{amsfonts}
\usepackage[margin=1in]{geometry}

\title{The Concatenator: A Bayesian Approach To Real Time Concatenative Musaicing}

\author{ Christopher J. Tralie \\ 
Ursinus College Department of Mathematics, Computer Science, And Statistics \\ 
\tt ctralie@alumni.princeton.edu
\and 
Ben Cantil \\ 
DataMind Audio \\ 
\tt bencantil@gmail.com}

\begin{document}

\maketitle

\begin{abstract}
    We present ``The Concatenator,'' a real time system for audio-guided concatenative synthesis. Similarly to Driedger et al.'s ``musaicing'' (or ``audio mosaicing'') technique, we concatenate a set number of windows within a corpus of audio to re-create the harmonic and percussive aspects of a target audio stream. Unlike Driedger's NMF-based technique, however, we instead use an explicitly Bayesian point of view, where corpus window indices are hidden states and the target audio stream is an observation. We use a particle filter to infer the best hidden corpus states in real-time. Our transition model includes a tunable parameter to control the time-continuity of corpus grains, and our observation model allows users to prioritize how quickly windows change to match the target. Because the computational complexity of the system is independent of the corpus size, our system scales to corpora that are hours long, which is an important feature in the age of vast audio data collections. Within The Concatenator module itself, composers can vary grain length, fit to target, and pitch shift in real time while reacting to the sounds they hear, enabling them to rapidly iterate ideas. To conclude our work, we evaluate our system with extensive quantitative tests of the effects of parameters, as well as a qualitative evaluation with artistic insights. Based on the quality of the results, we believe the real-time capability unlocks new avenues for musical expression and control, suitable for live performance and modular synthesis integration, which furthermore represents an essential breakthrough in concatenative synthesis technology.
    
\end{abstract}

\section{Introduction}

Concatenative synthesis, or audio mosaicing, is a data-driven approach to arrange granular fragments of audio samples, particularly using data sourced from the spectral-temporal features of a target sound. While granular synthesis systems typically rely on combinations of aleatoric parameterization, deterministic automation, and traditional synthesis modulation to achieve complex and evolving textures from sound fragments \cite{roads1978granular}, concatenative synthesis algorithms utilize Music Information Retrieval technology to decide parameters such as the index, amplitude, and pitch of each sound fragment. 

Modern music producers are inundated by audio data. Services like Splice offer hundreds of thousands of samples readily available on the cloud, and Kontakt multi-sample libraries can often take up over 10gb of disk space to capture a single instrument. Music Producers generate plenty of their own audio data as well: stems, multi-tracks, long-form recordings, and mix variations account for a large portion of many a music producer's audio collection. Recent software such as XO by XLN Audio, Sononym, and Ableton Live 12 offer automatic organization of audio files based on various tags and descriptors, but these implementations of MIR technology are more utilitarian than creative in their design and application. Meanwhile, concatenative synthesis options remain sparse since its conceptual inception \cite{schwarz2000system}: Reformer by Krotos is designed to create foley designs, apps like Samplebrain and CataRT \cite{schwarz2006real, schwarz2008principles} are lacking in critical musical areas such as pitch tracking, with the more advanced options having limited accessibility for artists, requiring prior knowledge of Max (FluCoMa, MuBu) or Python (Audioguide).

The Concatenator advances concatenative synthesis in 3 major ways: 1) it is capable of accurately reproducing harmonic and percussive sounds using arbitrary corpora 2) in real-time at scale, 3) affording new levels of control and accessibility.  Furthermore, unlike neural audio systems \cite{bitton2020neural}, it requires no training and can adapt to arbitrary corpora at runtime. The speed, ease, and scope of The Concatenator offers a fresh paradigm for music producers to interact creatively with their ever-expanding excess of audio data, leading to what we believe is a breakthrough in the field.

\section{Related Work}
\label{sec:relatedwork}
We build on important works in Bayesian inference, particle filters, concatenative synthesis, and applied nonnegative matrix factorization (NMF), which we briefly describe

\textbf{Driedger's Technique.} From an artistic point of view, the most similar technique to ours is Driedger et al.'s 2015 ``Let It Bee'' concatenative musaicing technique \cite{driedger2015let}, which uses NMF to learn activations of spectral window templates in a \textbf{corpus collection} so that their combination will match a \textbf{target} spectrogram. This technique was a fruitful innovation in sound design for electronic music production, as featured heavily on {\em Zero Point} by Rob Clouth\cite{clouth2020}, using custom software also authored by Clouth. The algorithm was also implemented in an open source python script in 2018 \cite{tralie2018}, and in Max by the FluCoMa project in 2021 (fluid.bufnmfcross)\cite{flucoma2021}, which made NMF-inspired audio mosaicing accessible enough to contribute towards the production of at least two more albums heavily featuring the technique: {\em Edenic Mosaics} by Encanti (2021) \cite{cantil2021} and {\em Hate Devours Its Host} by Valance Drakes (2023) \cite{drakes2023}.

We now detail the mathematics of Driedger et al.'s technique, as we borrow a few ideas in our work.  Driedger et al. learn $H$ in the equation $V \approx WH$, where $V$ is an $M \times T$ target spectrogram with $M$ frequency bins and $T$ times, $W$ is an $M \times N$ set of $N$ spectral corpus templates that are treated as fixed, and $H$ is a matrix of $N \times T$ learned activations.  For instance, $W$ could be the windows of a collection of buzzing bees and $V$ could be an excerpt from The Beatles' ``Let It Be'' (hence the title).  Driedger et al. use the Kullback-Liebler (KL) divergence loss, an instance of the more general $\beta$-divergence \cite{buch2017nichtnegativematrixfaktorisierungnutzendesklangsynthesensystem}, to measure the goodness of fit of $WH$ to $V$.  This loss function is 

\begin{equation}
\label{eq:klloss}
D(V || WH) = \sum V \odot \log \left( \frac{V}{WH} \right) - V + WH
\end{equation}

where $\odot$, $/$, $+$, and $-$ are all applied element-wise, and the sum is taken over all elements of the resulting matrix.  As Lee/Seung show, choosing the right step size turns gradient descent of Equation~\ref{eq:klloss}, with respect to $W$ and $H$, into {\em multiplicative update rules} that guarantee monotonic convergence.  Driedger et al. keep $W$ fixed to force the final audio to use exact copies of the templates, so only the update rule for $H$ is relevant.  At iteration $\ell$, this is:

\begin{equation}
\label{eq:klhgrad}
H_{kt}^{\ell} \gets H_{kt}^{\ell-1} \left( \frac{ \sum_{m} W_{mk} V_{mt} / (WH^{\ell-1})_{mt} }{ \sum_{m} W_{mk} } \right)
\end{equation}

Crucially, though, Driedger et al. note that the update rules in Equation~\ref{eq:klhgrad} alone will lose the timbral character of the templates in $W$.  They hence disrupt ordinary KL gradient descent by performing several increasingly impactful modifications to $H$ before Equation~\ref{eq:klhgrad} in each step, which are eventually set in stone after $L$ total iterations.  First, they avoid repeated windows to avoid a ``jittering'' effect, allowing a particular window $k$ to only activate once in some $r$-length interval based on where it's the strongest:
\begin{equation}
    \label{eq:driedgerrepeated}
    (H_r)_{kt}^{\ell} \gets \left\{ \begin{array}{cc} H^{\ell-1}_{kt} & H^{\ell-1}_{kt} > H^{\ell-1}_{ks}, |t - s| \leq r \\ H^{\ell-1}_{kt} (1 - \frac{\ell+1}{L}) & \text{otherwise}  \end{array} \right\}
\end{equation}
They also promote sparsity similarly by shrinking all but the top $p$ activations in each column of $H_r$ to create $H_p^{\ell}$.  Finally, they encourage {\em time continuous activations} by doing ``diagonal enhancement,'' or by doing a windowed sum down each diagonal of $H_p$, assuming the columns of $W$ are also in a time order.

\begin{equation}
    \label{eq:driedgertimecontinuous}
    (H_c)_{kt}^{\ell} = \sum_{i=-c}^c (H_p)^{\ell}_{k+i, t+i}
\end{equation}

Since this encourages the algorithm to mash up chunks of $W$ in a time order, it effectively encourages sound grains from the templates than the length of a single window that ordinary NMF would take.  Finally, Driedger et al. apply Equation~\ref{eq:klhgrad} to $H_c^{\ell}$ instead of $H^{\ell-1}$ to obtain $H^{\ell}$.

These disruptions remove the guarantee that Equation~\ref{eq:klloss} will be minimized, or that it will even monotonically decrease, but Driedger et al.'s key insight is that the loss function is merely a guide to choose reasonable activations; a suboptimal fit leaves room to better preserve timbral characteristics.  We take a similar perspective.

\textbf{Driedger Tweaks.} The idea of spectrogram decomposition used for concatenative musaicing goes back to the work of \cite{burred2013cross}.  Beyond that, the authors of \cite{buch2017nichtnegativematrixfaktorisierungnutzendesklangsynthesensystem} provide some improvements to Driedger et al.'s technique, including mixing corpus windows directly rather than performing phase retrieval on $W H$. One issue with Driedger et al.'s technique is the sources have to be augmented with pitch shifts to span additional pitches in the target, increasing memory consumption and runtime.  The authors of \cite{foroughmand2017multi, aarabi2018music} avoid this by using 2D deconvolutional NMF \cite{schmidt2006nonnegative} on the Constant-Q transform, whereby pitch shifts are modeled as constant shifts of the activations instead of the templates, saving memory.  The other convoluational axis models time history and time shifts, avoiding the need for the diagonal enhancement of Equation~\ref{eq:driedgertimecontinuous}.  The authors apply 2D NMF to both the source and target, so they do not preserve the original sound grains.  However, for our preferred style, we want to take the source grains exactly as they are.

\textbf{Other Concatenative Techniques.} Schwarz created an offline concatenative synthesis system dubbed ``Caterpillar'' that uses the Viterbi algorithm \cite{schwarz2000system}, which he later approximated with a real time system, ``CataRT'' that uses a greedy approach instead of the Viterbi algorithm \cite{schwarz2006real, schwarz2008principles}.  Simon's ``audio analogies'' is quite similar \cite{simon2005audio}, but instead of a user controlled traversal through timbral space, they use features from some source (e.g. midi audio) to guide synthesis to a target with a different timbre (e.g. real audio of someone playing a trumpet). Caterpillar and audio analogies are both {\em sequentially Bayesian in nature}, where the {\em hidden state} is the template to concatenate, and the ``observation'' is a user-controlled trajectory or features from a source timbre, respectively.  The prior transition probabilities are based on temporal continuity.  However, they use the Viterbi algorithm, which is computationally intensive and which needs all time history, so it cannot be applied in real time.  By contrast, a \textbf{particle filter} is a scalable Monte Carlo method for sequential Bayesian inference \cite{metropolis1949monte, doucet2000sequential, thrun2002probabilistic}.  It is less common in MIR, but it has found use in a few real time MIR applications such as multi-pitch tracking \cite{duan2011state}, tempo tracking \cite{cemgil2003monte, hainsworth2004particle}, and beat tracking \cite{heydari2021don}.  

\begin{figure*}[h]
	\centering
	\includegraphics[width=\textwidth]{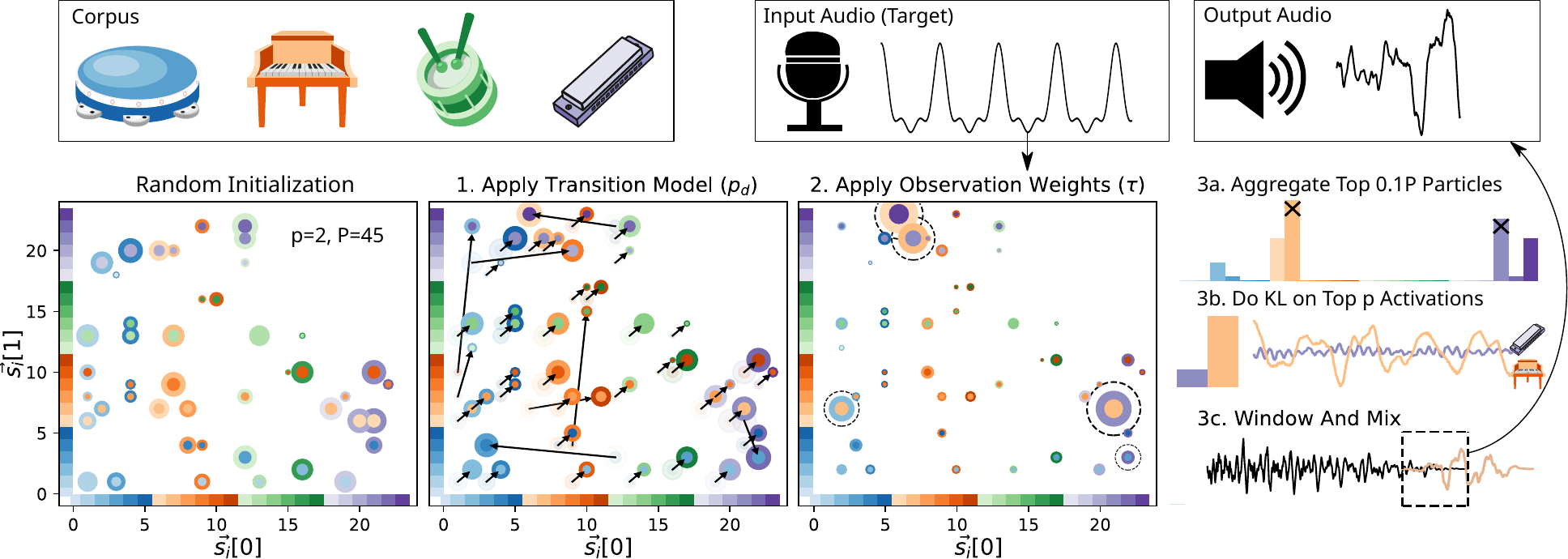}%
	\caption{\textbf{The Concatenator} maintains $P$ ``particles,'' each of which represents $p$ specific corpus windows.  Each window moves forward by 1 timestep in the corpus with probability $p_d$, or otherwise jumps randomly.  Then, particles each mix their windows to best match the target, and particles with the top 10\% best fits to the target vote on a final set of $p$ windows.}
	\label{fig:BlockDGM}
\end{figure*}

\section{The Concatenator}

The NMF technique of Driedger is not suitable for real-time applications; the gradient update rules of Equation~\ref{eq:klhgrad} scale linearly in the length of the corpus, leaving all but minutes long corpora usable (Section~\ref{sec:complexity}), and the equations to suppress repeated windows and promote time continuity at each entry of $H$ require knowledge of all activations in $H$, including future activations.  Instead, The Concatenator does many tiny KL-based NMF problems (Equation~\ref{eq:klhgrad}) online in ``particles'' based on random sampling at each timestep.  The particles then vote on a final set of activations to use at that timestep (Figure~\ref{fig:BlockDGM})\footnote{CC musical instrument images adapted from \url{vectorportal.com}}.  The random sampling trades off historical context to choose longer grains, with fit to the target audio streaming in.  We provide the mathematical and implementation specifics below.

\subsection{Sequential Bayesian Formulation And State Space}
\label{sec:bayesian}

    Formally, The Concatenator uses a sequential Bayesian formulation, where the $t^{\text{th}}$ column of the target spectrogram $V$ is the ``observation,'' at time $t$.  The hidden state indexes $p$ out of $N$ possible windows in the corpus spectrogram $W$.  We use a particle filter to efficiently infer the the best such windows (Section~\ref{sec:sampling}). Henceforth, we refer to the observations as vectors $\vec{v}_t$ to emphasize that the data is streaming, and we focus on one timestep $t$ at a time.  
    
    \textbf{State space.} To keep the state space simple, we decouple which windows are active from their activation weights; we only model the former as the hidden state, while we infer the weights as a best fit under the KL-loss (Equation~\ref{eq:klloss}).  To control for polyphony directly, we use a {\em $p$-sparse} nonnegative integer-valued vector $\vec{s}_t \in \mathbb{N}^{p}$ as the hidden state.  This vector indexes the $p$ corpus windows that are active at time $t$, where $p$ is fixed ahead of time.  For convenience of implementation, template indices can repeat and are in no particular order:\footnote{One could also model $\vec{s}_t$ as a \textbf{set} of $p$ elements, which would identify all $\vec{s}_t$'s that are permutations of each other.  This would reduce the cardinality of the state space to $\binom{N}{k}$ at the cost of a more cumbersome implementation.  We also rely on some redundancy sampling good activations in our particle filter, so allowing repetitions does not hurt performance.}:

    \begin{equation} 
        \label{eq:statevector}
        \vec{s_t}[k] \in \left\{0, 1, ..., N-1\right\}, k = 0, 1, ..., p-1 
    \end{equation}

    We then infer the associated nonnegative weights $\vec{h_t}[k]$ for each activation to give the approximation $\vec{\Lambda_t}$ at time $t$:
    
    \begin{equation}
        \label{eq:approximation}
        \vec{\Lambda_t}[m] = \sum_{k=0}^{p-1} \vec{h_t}[k]  W_{m, \vec{s_t}[k]}
    \end{equation}

    In particular, given $W$, $\vec{s_t}$, and $\vec{v_t}$, we apply the update rules of Equation~\ref{eq:klhgrad} for a pre-specified number $L$ of iterations, using the corresponding columns $\vec{s_t}$ of $W$

    \begin{equation}
        \label{eq:sparseklhgrad}
        \vec{h_t}^{\ell}[k]  \gets \vec{h_t}^{\ell-1}[k] \left(  \frac{\sum_m (W_{m, \vec{s_t}[k]}) (\vec{v_t}[m]) / (\vec{\Lambda_t}^{\ell-1}[m]) }{ \sum_{m} W_{m, \vec{s_t}[k]}} \right)
    \end{equation}

    \textbf{Transition Model.} We use the KL-loss (Equation~\ref{eq:klloss}) to measure the spectral fit of $\Lambda_t$ to $\vec{v_t}$.  As in Driedger et al. \cite{driedger2015let} (Equation~\ref{eq:driedgertimecontinuous}), however, we are willing to sacrifice fit to take longer grains from the corpus $W$.  To that end, we define the prior  \textbf{state transition probability} in the as a Factorial Hidden Markov Model (FHMM) \cite{ghahramani1995factorial}. Each $\vec{s_t}$ satisfies the Markov property and is conditionally independent of all previous steps given $\vec{s_{t-1}}$, but {\em each component $k$} of $\vec{s_t}[k]$ also transitions independently of other components, leading to the following transition probability:

    \begin{equation}
        \label{eq:transitionprob}
        p_T(\vec{s_t} = \vec{b} | \vec{s_{t-1}} = \vec{a}) = \prod_{k=0}^{p-1} \left\{  \begin{array}{cc}  p_d & \vec{b}[k] = \vec{a}[k]+1  \\ \frac{1-p_d}{N-1} & \text{otherwise} \end{array} \right\}
    \end{equation}
    where $p_d \in [0, 1]$ is the ``probability of remaining time-continuous.''  Intuitively, if $p_d > 0.5$, then we are more likely to continue to use a time-continuous activation than we are to jump to a new random activation, which promotes longer contiguous sound grains from the corpus, even at the expense of a lower fit to the spectral template\footnote{This has a similar effect to ``extend matches'' functionality in Sturm's MatConcat \cite{sturm2004matconcat} when a match isn't found. In our Bayesian framework, such extensions happen on a continuum based on fit to target.}.  As such, $p_d$ a parameter that can be tuned by the artist and set closer to $1$ to promote longer grains.  We generally find $p_d \in [0.9, 0.99]$ to be effective (Section~\ref{sec:quantitativeevaluation}).

    We must also specify the {\em observation probability}, which pulls the states closer to matching $\vec{v_t}$, even if they have to jump away from time continuity; otherwise, the result would sound nothing like the target.  Though each component transitions independently, they all contribute jointly to an observation, which makes inference trickier than it is for traditional HMMs.

\subsection{Sampling, Observing, And Synthesizing}
\label{sec:sampling}

We now describe how to apply Bayesian inference to find the sequence of corpus windows $\vec{s_t}$ and their activation weights $\vec{h_t}$ that maximize the posterior probability given the transition model in Equation~\ref{eq:transitionprob} and the observation model below.  While the authors of \cite{wohlmayr2010probabilistic} use a similar FHMM applied to multi-pitch tracking, inferring the hidden states via message passing algorithms known as ``Max-Sum'' \cite{kschischang2001factor} and ``Junction Tree'' \cite{jensen1996introduction}, we need a faster technique which is also real-time, and which has tunable accuracy that degrades gracefully with restricted computational resources.  To that end, we turn to a particle filter.

Our particle filter consists of $P$ particles, each of which is a $p$-dimensional state vector (Equation~\ref{eq:statevector}) that we refer to as $\vec{s_i}$.  The particles traverse the corpus over time, and they each have a weight $w_i$ that keeps track of the posterior probability of its accumulated motion over all timesteps (we now dispense with the time index $t$ on $\vec{s_i}$ and $w_i$ since $t$ will be clear from context).  Since each particle is its own estimate of a state that best describes what templates to choose, our goal is to sample them in such a way that (at least some of) the particles are close to capturing activations that maximize the posterior probability given all $\vec{v_t}$.

\textbf{Tracking Weights.} All particles begin with even weights $w_i = 1/P$.  At the beginning of each time step, we sample new indices for each $\vec{s_i}$ according to Equation~\ref{eq:transitionprob}.  Then, we multiply each weight by the \textbf{observation probability $p_O$}.  Given the KL loss $D_i$ between the $i^{\text{th}}$ particle's spectral approximation $\vec{\Lambda_i}$ (Equation~\ref{eq:approximation}) and $\vec{v}_t$ after $L$ iterations of Equation~\ref{eq:sparseklhgrad}, for each particle $i$, $p_O$ is:

\begin{equation}
    \label{eq:observationprob}
    p_O[i] = \frac{e^{-\tau D_i}}{ \sum_{j} e^{-\tau D_j}}
\end{equation}

In other words, the observation probability is a softmax over KL-based goodness of fits of $\vec{s_i}$ to $\vec{v_t}$, and the softmax has a ``temperature'' $\tau$.  We use a negative exponential since a larger $D_i$ loss indicates a poorer fit using windows $\vec{s_i}$ and hence, should be a lower probability.  Intuitively, a higher $\tau$ will emphasize particles that fit the observation better, putting more importance on the observation relative than the transition probability.  This is tunable and has a similar effect to varying $p_d$ in the transition, as we will explore more in Section~\ref{sec:quantitativeevaluation}.  After multiplying each $w_i$ by $p_O[i]$, we normalize the weights so that they sum to 1.  
If the sum of the weights is too numerically small based on the chosen floating point precision, we reset each weight to be $1/N$.

\textbf{Resampling.} The above is a naive particle filter, but it suffers from ``sample impoverishment,'' where a few particles stand out with high weights and the rest are stuck with vanishing weights, leaving the system unable to adapt to new observations.  To ameliorate this, we compute a standard definition of the ``effective number of particles'' $n_{\text{eff}} = 1/(\sum_{i} w_i^2)$, which is maximized when all particles have equal weight $1/P$.  If $n_{\text{eff}}$ goes below $0.1P$ at a particular time step, we resample the particles with stochastic universal sampling \cite{kitagawa1996monte, carpenter1999improved}, an $O(P)$ resampling technique, and reset all weights to $1/P$ before continuing.  This leads to ``survival of the fittest'' where particles with a higher weight are more likely to be replicated and those with a lower weight are more likely to be eliminated.

\textbf{Synthesizing audio.} After updating the weights, we take a weighted average of the windows in the top $0.1P$ particles, with the option to further boost windows that follow continuously from those chosen in previous steps.  We also ignore windows that would be repeated from up to $r$ timesteps in the past (analogous to Driedger's Equation~\ref{eq:driedgerrepeated}).  We then let $\vec{s_t}$ be the top $p$ such windows by weight, and we compute the corresponding activations $\vec{h_t}$. These steps can be done in $O(Pp)$ time with hash tables and linear time selection.  Finally, we mix together the corresponding waveforms from the corpus (as in \cite{buch2017nichtnegativematrixfaktorisierungnutzendesklangsynthesensystem}) and apply a Hann window to overlap-add this audio to the output stream.

\subsection{Computational Complexity} 
\label{sec:complexity}
The dominant cost of both The Concatenator and of Driedger is computing activations via KL iterations.  Given $N$ corpus templates, $T$ times in the target, and a spectral dimension of $M$, for $L$ KL iterations, the time complexity of Driedger (Equation~\ref{eq:klhgrad}) is $O(LMNT)$.  This is a {\em linear} dependency on the corpus length.  So if, for example, Driedger's technique takes a minute on a target sourcing a corpus that's a minute in length, it will take 2 hours a 2-hour corpus on that same target.  To improve this scaling, the authors of \cite{buch2017nichtnegativematrixfaktorisierungnutzendesklangsynthesensystem} do a greedy nearest neighbors search in the corpus, but this requires tuning and may miss important windows.  In fact, our random sampling naturally scales in an even more favorable way.  Specifically, given $P$ particles and $p$ windows per particle, the time complexity of our analogous Equation~\ref{eq:sparseklhgrad} is only $O(LPMpT)$, {\em which does not scale with the corpus size $N$  at all} (though $P$ may need to scale with $N$ for the best results (Section~\ref{sec:quantitativeevaluation})).  As an example, for a 60 minute corpus a window length of 2048 ($M=1025$, hop=$1024$) at a sample rate of 44.1khz, using $P=1000$ and $p=5$, this is a speedup of nearly 30x over Driedger.  Moreover, propagating particles and applying the observation model are also embarrassingly parallelizable at the particle level, which we leverage in our implementation.  Finally, while Driedger et al. use $L=20$ \cite{driedger2015let}, we find that $L=10$ is sufficient in our context.

\subsection{Bells And Whistles (Pun Intended)}

\textbf{Regularizing Quiet Moments in The Corpus.} One pitfall using KL-based NMF is that if enough activations are near silence, Equation~\ref{eq:sparseklhgrad} becomes numerically unstable and the weights $\vec{h_i}$ can approach $\infty$.  To address this, we modify the KL-loss to include a masked $L_2$ penalty for $\vec{h_i}$ for the $i^{\text{th}}$ particle for the target $\vec{v_t}$ at time $t$.  Given the corresponding approximation $\vec{\Lambda_i}$ (Equation~\ref{eq:approximation}), the modified loss is

\begin{equation}
    \label{eq:kllossalpha}
    D_i(\vec{v_t} || \vec{\Lambda_i}) = \left( \sum \vec{v_t} \odot \log \left( \frac{\vec{v_t}}{\vec{\Lambda_i}} \right) - \vec{v_t} + \vec{\Lambda_i} \right) + \boldsymbol{ \frac{||\alpha \odot \vec{h_i}||_2^2 }{2} }
    \end{equation}

where, abusing notation, $\alpha$ is a mask that is a fixed value (we use 0.1) if the corresponding corpus window is less than -50dB and $0$ otherwise.  Equation~\ref{eq:sparseklhgrad} then turns into 

\begin{equation}
    \label{eq:sparseklhgradalpha}
    \vec{h_i}^{\ell}[k]  \gets \vec{h_i}^{\ell-1}[k] \left(  \frac{\sum_m (W_{m, \vec{s_i}[k]}) (\vec{v_t}[m]) / (\vec{\Lambda_i}^{\ell-1}[m]) }{ (\sum_{m} W_{m, \vec{s_i}[k]}) + \boldsymbol{ \alpha[k] \vec{h_i}^{\ell-1}[k]}} \right)
\end{equation}

Intuitively, if $s_i[k]$ is a quiet corpus window, $\alpha[k] = 0.1$, which shrinks $\vec{h_i}^{\ell-1}[k]$ down\footnote{For a derivation of similar additive constraints on NMF, refer to \cite{virtanen2007monaural}}.

\textbf{Mel Spectrograms.} The computational complexity scales linearly with $P$, so lowering $P$ makes the system run faster, but it is also possible to decrease $M$ without sacrificing much quality.  One can replace the full spectrogram with a mel-spaced spectrogram with many fewer bins.  In fact, Schwarz's Caterpillar system did something very similar, suggesting alternative features to the raw spectrogram such as spectral flux \cite{schwarz2000system}.

\textbf{Pitch Shifting.} Though we don't use this in Section~\ref{sec:evaluation}, we implemented Driedger et al.'s technique to increase the pitch coverage of the corpus; that is, we can replicate the corpus in its entirety for different pitch shifts that are chosen up-front.  This only incurs a preprocessing cost since the complexity of The Concatenator is independent of corpus length (Section~\ref{sec:complexity}), which does not impact real-time performance once the system starts.  However, our system could choose a different trade-off of space and time complexity by augmenting the state space as the Cartesian product of window indices and pitch shifts.  Pitch shifts could be computed on the corpus audio {\em on demand} whenever a state with a nonzero pitch shift is chosen.

Finally, for a fixed corpus with or without pitch shifts, the user can control a slider that pitch shifts the {\em target} in real time, so that the chosen windows move relatively to the audio input.  This could be used, for example, to harmonize to singing in an interval that's a fifth away.

\subsection{How Many Particles?}
\label{sec:howmanyparticles}
In practice, few particles are surprisingly effective at capturing windows that fit the target, which we explain with a simple probabilistic argument.  Given a corpus with $N$ sound grains (including pitch shifts) and $P$ particles that each capture $p$ windows, suppose also that we have a hypothetical ``ideal particle'' $\vec{s_t}$ with the $p$ best windows at time $t$, which are completely disjoint from all current particles; the only way to jump to the best windows is to randomly resample with probability $(1-p_d)$.  Since we use a small hop length relative to the sample rate (1024/44100 $\approx$ 23 ms), we have a few timesteps to jump without a large effect on the final audio.  Also, there are usually several windows in the corpus that sound acceptably similar to windows in $\vec{s_t}$.  Let $\delta$ be the maximum tolerable offset before or after in time for choosing the best windows, and let $w$ be a factor of acceptable windows (e.g. $w=11$ would consider each window in $\vec{x}$ and its ten most similar in the corpus).  Assuming all offsets of acceptable windows are disjoint, then the probability of jumping to at least one of the top $k$ windows of $\vec{s_{t}}$, or to one of their acceptably close corresponding offsets, is:

\begin{equation}
    \label{eq:timeadjacentprobmodified}
    1 - \left( p_d + (1-p_d) \frac{(N-1-wk)}{N-1} \right)^{(2 \delta +1)pP}
\end{equation}

For example, for $p_d = 0.95$, $\delta=2$ and $w = 11$, and $N=10000$ ($\approx$ 4min corpus), the probabilities are 0.747, 0.936, 0.983 for $k=1, 2, 3$, respectively.  These probabilities all degrade when $N$ gets larger for a larger corpus, but in that case, it is likely that the acceptable $w$ is also larger.

Furthermore, once one of the particles catches on to a good window in the corpus, it is promoted with a high weight and gets carried on to a longer grain.  This is similar to how the ``patch match'' technique in computer graphics \cite{Barnes:2009:PAR, Barnes:2010:TGP} computes nearest neighbors of many nearby patches by starting with a random initialization of nearest neighbors, and then well-matched to patches correct the nearest neighbors of  spatially adjacent patches \cite{Barnes:2009:PAR}.

\begin{figure}
    \centering
    \includegraphics[width=0.7\columnwidth]{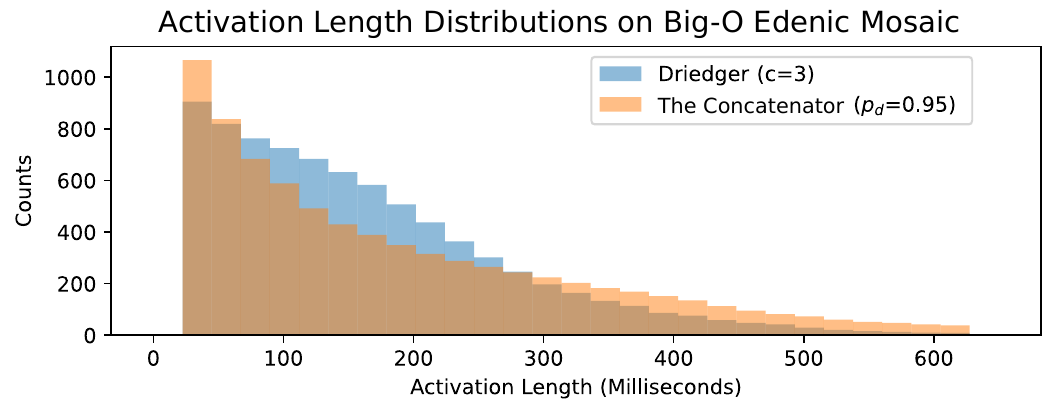}
    \caption{The activation distribution of The Concatenator is more exponential in nature than that of Driedger et al.  In fact, it exactly follows the geometric distribution if the tempreature $\tau = 0$.}
    \label{fig:ActivationsDist}
\end{figure}

\textbf{Grain Lengths.} In the absence of observation re-weighting and re-sampling, Equation~\ref{eq:transitionprob} leads to a geometric distribution with a mean grain length of $\text{sr}/(\text{hop}(1-p_d))$.  Even with observation probabilities shortening the lengths, the grain lengths are still ``exponential'' in character; there are more very short activations and more very long activations than Driedger's technique, even for similar means (Figure~\ref{fig:ActivationsDist}). This leads to subtle auditory differences.

\section{Evaluation}
\label{sec:evaluation}

\subsection{Quantitative Evaluation}
\label{sec:quantitativeevaluation}

To empirically assess reliability, we do an extensive MIR-style evaluation, which is much more comprehensive than standard evaluation in other concatenative synthesis works.

\textbf{Effect of Parameters.} First, to complement our analysis in Section~\ref{sec:howmanyparticles}, we want to empirically examine how many particles are needed for different sized corpora.  We also want guarantee the impact of important parameters in our system for artistic control.  We select 3 corpora: Driedger's buzzing bees (small, 66 seconds), a corpus used in {\em Edenic Mosaics}\cite{cantil2021} known as ``EdenVIP2,'' which consists of various real-world percussive sounds (medium, ~10.5 minutes), and all Woodwind clips from the pre-2012 UIowa MIS dataset \cite{uiowadataset} (large, $\approx$1.6 hours).  Then, we randomly subsample 1000 30 second clips from the Free Music Archive (FMA)-small dataset \cite{fma_dataset}, each of which we use as a target for the three different corpora for various parameter choices.  We use a sample rate of 44.1khz for all corpora, we use stereo audio for the bees and EdenVIP2, and we use mono audio for the Woodwinds.

\begin{figure}
    \centering
    \includegraphics[width=\columnwidth]{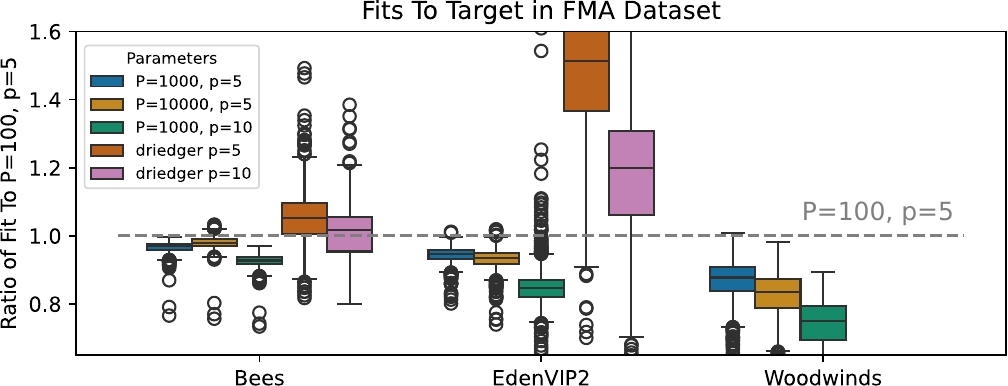}
    \caption{Increasing polyphony leads to a better fit (ratios $<1$), and increasing particles leads to a better fit, especially for larger corpora like the Woodwinds ($\approx$1.6hrs).}
    \label{fig:fmafit}
\end{figure}

First, we assess the effect of particles on fit; we fix $p_d=0.95$, temperature $\tau=10$, and $r=3$, using $L=10$ iterations for all KL operations, and we take $P \in \left\{ 100, 1000, 10000 \right\}$.  We also compare to Driedger et al.'s technique with $c=3$ and $r=3$ using $L=50$ iterations, though we omit comparisons with Woodwinds due to computational cost (Section~\ref{sec:complexity}).  In all cases, we use frequencies from $0$ to $8000$hz with a sample rate of 44100hz, a window length of 2048 samples, and a hop length of 1024 samples.  Since the spectral similarity of different targets to a particular corpus varies widely, we report the {\em ratio} of the KL loss in Equation~\ref{eq:klloss} to the KL loss for The Concatenator with $P=100, p=5$.  Figure~\ref{fig:fmafit} shows the results.  As expected, an increased polyphony leads to a better fit, as does increasing particles for all but the Bees, though the effect of increased particles is most pronounced for the largest corpus of Woodwinds, which makes sense by Equation~\ref{eq:timeadjacentprobmodified}.

\begin{figure}
    \centering
    \includegraphics[width=\columnwidth]{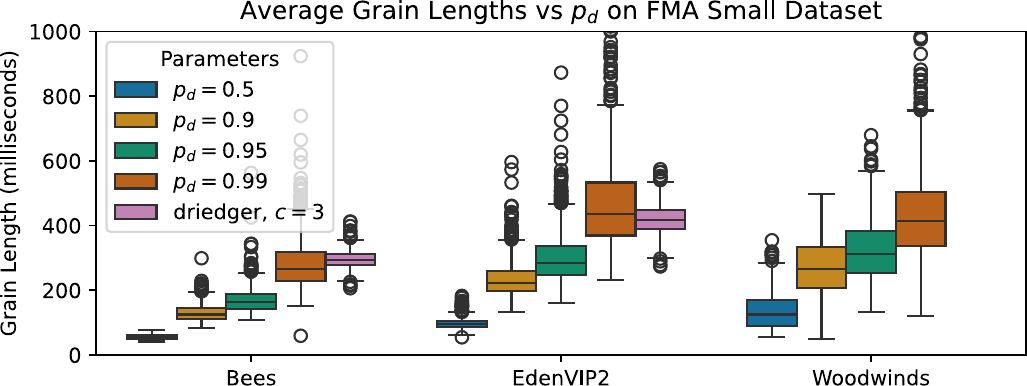}
    \caption{Increasing $p_d$ increases the average grain length since windows are less likely to jump at each timestep.}
    \label{fig:pdGrainLengths}
\end{figure}

\begin{figure}
    \centering
    \includegraphics[width=\columnwidth]{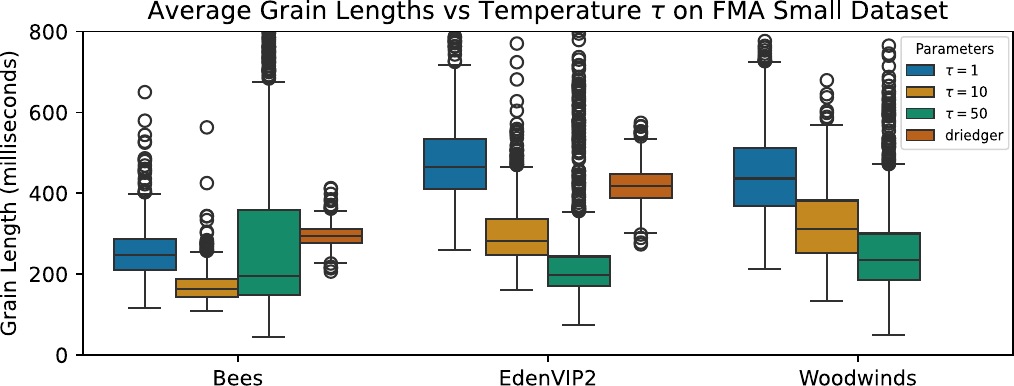}
    \caption{Increasing $\tau$ {\em decreases} the average grain length since this prioritizes the observation probability.}
    \label{fig:tempGrainLengths}
\end{figure}

As we noted in Section~\ref{sec:relatedwork}, however, a very good fit may lose the timbral characteristics of the corpus.  A lower $p$ helps, but we also need to ensure that grains are long enough.  Therefore, we also examine mean grain length for various parameters.  Figure~\ref{fig:pdGrainLengths} shows the result of varying $p_d$ for a fixed temperature $\tau=10$ and $p=5$, and Figure~\ref{fig:tempGrainLengths} shows the result of varying the temperature $\tau$ for $p=5$ and $p_d = 0.95$.  As expected, grain length goes up with increased $p_d$ and down with increased $\tau$.  In practice, lowering $\tau$ and raising $p_d$ will lead to especially long grain lengths, albeit with a lower target fit.

\begin{figure}
    \centering
    \includegraphics[width=\columnwidth]{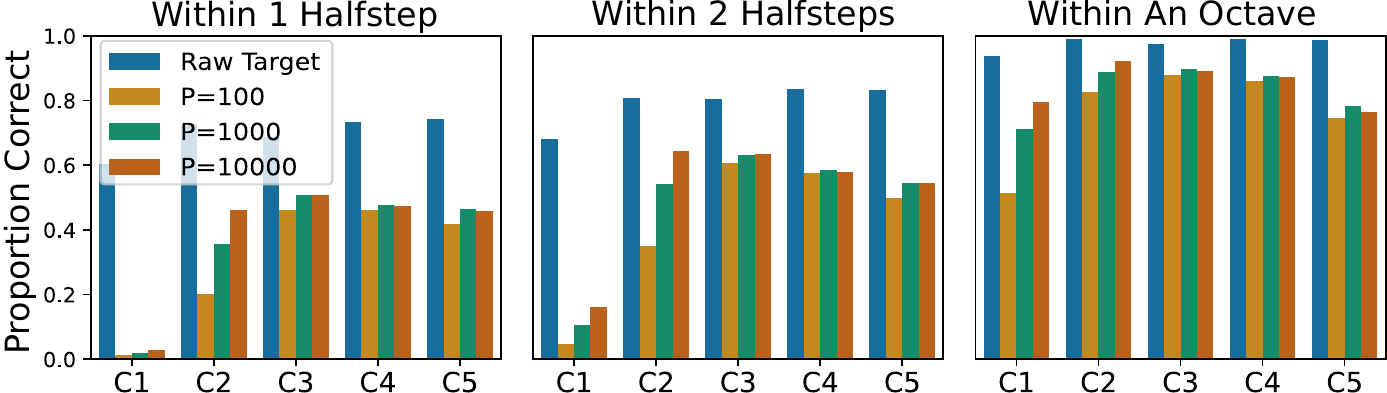}
    \caption{The Concatenator reproduces reasonably correct pitches in the 1.6 hour Woodwinds corpus with targets in MDB-stem-synth, in real time (at $P=100, 1000$), at all but the lowest octave C1.}
    \label{fig:PitchTests}
\end{figure}

\textbf{Reproducing Pitch.} In addition to fits and grain lengths, we quantify how well The Concatenator reproduces target pitch.  Using the Woodwinds corpus, we create targets out of all stems in the MDB-stem-synth dataset \cite{salamon2017analysis}.  We compare ground truth pitch annotations of the stems to the pitches estimated with CREPE \cite{kim2018crepe} on both the raw target and the synthesized audio for various $P$, and we break the results down by octave.  Figure~\ref{fig:PitchTests} reports the proportion of pitches correctly identified at each 23ms hop length to within different tolerances, over all stems.  Even though CREPE was not trained on concatenated audio, it reports pitch nearly as clearly as on the raw target for most octaves except for C1, which makes sense since the spectral resolution is only 21.5hz.  We can mitigate this in the current system by increasing the window, at the expense of temporal resolution.  In the future, though, we would like to try a streaming CQT that can better capture lower frequencies.  Finally, since the bassoon is the only instrument out of 10 in the Woodwinds that has notes in the C2 octave, additional particles are needed for precise pitch in that octave, which can be explained by $w$ in Equation~\ref{eq:timeadjacentprobmodified}.

\subsection{Qualitative Evaluation}
\label{sec:qualitative}

This algorithm was tested in a variety of contexts to assess its performance and accuracy for applications in music and sound design. Our Corpora contained audio samples that fell into the following categories: Test Tones, Percussion, Full Mixes, Sample Libraries, Foley, and Driedger Comparisons. Our Targets were single audio files that were designed to test how the Concatenator re-created varying kinds of melody, counterpoint, full mixes, basses, drums, vocals, noise, and prior examples used with the Driedger algorithm. 
Our tests reveal that the Concatenator performs highly accurately in pitch reproduction for most melodies, two-part harmonies, and full mixes that contain prominent melodic features, while struggling with accurate reproduction of more complex three-part harmony. 
Given the nature of the particle filter, which rotates through new temporal positions in the corpus at random, some notes are more accurate than others, and some notes are dropped all together, as expected from our quantitative analysis. While this tendency might make the Concatenator unfit for replacing the role of large multi-sample instruments, the vast majority of pitches remain wholly accurate while the aleatoric variation of off-color audio grains may represent an entirely desirable aesthetic quality of its own.
Similarly for drums, sometimes transients are incredibly accurate, while other times they sound a little smeared. This tendency is due to the particle filter's random positioning, and can be improved by increasing the particle amount.

\subsection{Supplementary Material / Discussion}

We include supplementary material at \url{https://www.ctralie.com/TheConcatenator}.  This includes a python prototype for the real-time system that uses port audio \cite{bencina2001portaudio}, audio examples for all corpus/target pairings in Section~\ref{sec:qualitative}, and a video showing artistic examples of what the real time system enables in the loop with Ableton Live.

This is only the beginning. Since The Concatenator exists feedback loop, we expect artists will go much deeper, possibly well beyond the ``obstacle course'' we put it through.

\bibliographystyle{plain}
\bibliography{main}

\end{document}